\PassOptionsToPackage{table}{xcolor}

\documentclass{aa}  

\usepackage{graphicx}
\usepackage{mwe}
\usepackage{longtable}
\usepackage{siunitx}
\usepackage{indentfirst}
\usepackage{orcidlink}
\usepackage{chemist}
\usepackage{multirow}
\usepackage{caption}
\usepackage[flushleft]{threeparttable}
\usepackage{subcaption}
\usepackage{float}
\usepackage{txfonts}
\titlerunning{}

\usepackage{natbib,twoopt}

\bibpunct{(}{)}{;}{a}{}{,}  

\makeatletter
  \newcommandtwoopt{\citeads}[3][][]{\href{http://adsabs.harvard.edu/abs/#3}%
    {\def\hyper@linkstart##1##2{}%
     \let\hyper@linkend\@empty\citealp[#1][#2]{#3}}}
  \newcommandtwoopt{\citepads}[3][][]{\href{http://adsabs.harvard.edu/abs/#3}%
    {\def\hyper@linkstart##1##2{}%
     \let\hyper@linkend\@empty\citep[#1][#2]{#3}}}
  \newcommandtwoopt{\citetads}[3][][]{\href{http://adsabs.harvard.edu/abs/#3}%
    {\def\hyper@linkstart##1##2{}%
     \let\hyper@linkend\@empty\citet[#1][#2]{#3}}}
  \newcommandtwoopt{\citeyearads}[3][][]%
    {\href{http://adsabs.harvard.edu/abs/#3}
    {\def\hyper@linkstart##1##2{}%
     \let\hyper@linkend\@empty\citeyear[#1][#2]{#3}}}
\makeatother

\usepackage{color}

\newcommand{\kms}{km\,s$^{-1}$}
\usepackage{cleveref}
\usepackage{chemist}

\AtBeginDocument{
\hypersetup{
    colorlinks=true, pdfstartview=FitV, 
    linkcolor=blue, citecolor=blue, 
    urlcolor=blue, breaklinks=true
}}

\begin{document}
\title{Discovery of ionized circumstellar gas emission around the long-period Cepheid $\ell$~Carinae with ALMA}

\titlerunning{Ionized circumstellar emission around the long-period Cepheid $\ell$~Carinae with ALMA}
\authorrunning{Hocd\'e et al. }
   \author{V. Hocdé\inst{1}\orcidlink{0000-0002-3643-0366}
   \and T. Kamiński\inst{2}
   \and M. Lewis\inst{3}
   \and N. Nardetto\inst{4}
   \and P. Kervella\inst{5}
   \and G. Pietrzyński\inst{1}}
   \institute{Nicolaus Copernicus Astronomical Centre, Polish Academy of Sciences, Bartycka 18, 00-716 Warszawa, Poland\\
   email : \texttt{vhocde@camk.edu.pl}
   \and Nicolaus Copernicus Astronomical Center, Polish Academy of Sciences, Rabiańska 8, 87-100 Toruń, Poland
   \and Leiden Observatory, Leiden University, P.O. Box 9513, 2300 RA Leiden, The Netherlands
   \and Universit\'e Côte d'Azur, Observatoire de la C\^ote d'Azur, CNRS, Laboratoire Lagrange, France
   \and LESIA, Observatoire de Paris, Université PSL, CNRS,  Sorbonne Université, Université Paris-Cité, 5 Place Jules Janssen, 92195 Meudon, France}


\date{Received 2 December 2024 ; accepted 22 January 2025}

\abstract{Cepheid circumstellar emissions have previously been detected via both infrared excess and infrared interferometric observations at a few stellar radii. Those studies have shown that these circumstellar emission can be produced by ionized gas, however there is no direct observational evidence to confirm this hypothesis. In this letter we explore the continuum emission and a spectrum of the bright and long-period Cepheid, $\ell$~Car ($P=35.56\,$day) at millimeter-wavelengths in order to detect possible effects of ionized gas emission. We presented ALMA observations of $\ell$ Car in two spectral setups in  Band~6 (near 212 and 253\,GHz, respectively) and we compared the measured flux density to one expected for the stellar continuum. We also derived the spectral index and probed the presence of Radio Recombination Lines (RRL). We report statistically significant emission of about 3.5$\,$mJy in the two spectral ranges, which is about 2.5 times the stellar continuum emission. For the first time, we are also able to derive the spectral index of the flux density ($S_\nu \propto \nu^\alpha$), $\alpha=+1.26\pm$0.44 ($\sim$3$\sigma$ error), which is characteristic of partially optically thick ionized gas emission. Additionally, we discovered an emission line from a RRL of hydrogen H29$\alpha$ centered on the stellar rest velocity, smaller in spatial extent than about 0\farcs2 ($\lesssim 100\,$AU), with a symmetric profile with a width at half power of 55.3$\pm$7.5\,\kms (1$\sigma$ error). It confirms the presence of ionized gas emission near $\ell$~Car. The millimeter emission detected from $\ell$ Car can be attributed to ionized gas emission from the Cepheid's chromosphere. Further radio interferometric observations are necessary to confirm the occurrence of these ionized gas envelopes around Cepheids of different pulsation periods.}

\maketitle

\section{Introduction}\label{s_Introduction}
The Cepheid $\ell$~Carinae ($\ell$~Car, HD 84810, HIP 47854) has the largest angular diameter among southern hemisphere Cepheids (in average 2.9$\,$mas), making it an ideal candidate for investigating its photosphere and circumstellar environment. Additionally, with a pulsation period of $P = 35.55\,$days, $\ell$~Car is the closest analog to the long-period, extragalactic Cepheids observed by the James Webb Space Telescope (JWST). Therefore, understanding the evolution and the physics of this star is important to infer the properties of the Period-Luminosity (PL) relation of long-period Cepheids, which directly impact the accuracy of the distance scale. This is especially important now, as JWST is observing Cepheids in distant galaxies to refine the precision and reliability of these measurements \citep{Freedman2024IAU, Riess2024}.

One of the intriguing phenomena associated with Cepheids is the possible presence of infrared (IR) circumstellar emission, which may serve as a tracer of their mass-loss rates \citep{Deasy1988} and may impact the accuracy of the PL relation \citep{Neilson2009,Neilson2010}. However, circumstellar emissions originate from various environments, including gas and dust contributions from the local interstellar medium, extended circumstellar material produced by the star's own mass loss, or an extended atmospheric layer near the stellar surface. Distinguishing between these components requires observations across different wavelengths and angular resolutions. For instance, \textit{Spitzer} IRAC imaging has revealed low surface brightness emission around $\ell$~Car at 5.8 and 8.0$\,\mu$m, extending out to a radius of approximately 35\arcsec (1.7 $\times 10^4$ AU; \citealt{Barmby2011}). Closer to the star, a comparative study in the IR with \textit{Spitzer}, VLT/VISIR, and VLTI/MIDI detected emission within a radius of 1\arcsec, attributed to warm circumstellar dust of a few 100\,K at a distance of about 500\,AU \citep{Kervella2009}. This region might be associated to the circumstellar emission resolved at only a few stellar radii by near-IR interferometry in the $K$-band through VLTI/VINCI observations around $\ell$~Car \citep{kervella06a}. Later, similar compact envelopes were resolved with CHARA/FLUOR \citep{merand06,merand07} around other Cepheids, as well as in the mid-IR, with MIDI and VLTI/MATISSE \citep{gallenne13b,Hocde2021}. However, evidence from \textit{Spitzer} low-resolution spectra and VLTI/MATISSE observations \citep{Hocde2020a,Hocde2025a} argue in favor of gaseous envelopes for most of the Cepheids, including $\ell$ Car. Therefore, these circumstellar emission could be related to an extended gas layer above the photosphere. Various authors reported a deep and static absorption feature in the H$\alpha$ line of $\ell$~Car \citep{Rodgers1968,nardetto08b,Hocde2020b}, and a strong Mg\,II absorption line in the ultraviolet (UV) \citep{love1994}, attributed to a gaseous envelope centered on the stellar restframe with an extension up to about 1000\,AU. Chromosphere activity is also suspected for several Cepheids, in particular for $\ell$ Car, from \ion{Ca}{II} K emission \citep{kraft57}, several UV emission lines \citep{schmidt1982I}, and the calcium IR triplet \citep{Hocde2020b}. The presence of hot plasma is also revealed by periodic UV emission lines and X-ray emission \citep[see, e.g.,][]{engle14,engle2017,Andrievsky2023}.
 On the modeling side, \cite{Hocde2020a} have shown that the IR excesses of Cepheids, including $\ell$ Car, can be explained by a thin envelope of ionized gas emitting free-free emission \citep{Hocde2021}. 
 \paragraph{}
 However, there is still no direct observational evidence of free-free emission toward Cepheids. Radio and millimeter-wave observations are useful to constrain ionized gas emission above the photospheric continuum as well as the spectral index. Prior to the recent results for $\delta$~Cep \citep{Matthews2023}, only upper limits on continuum emission were obtained for the twelve Cepheids observed with radio telescopes \citep{welch1988,Altenhoff1994}.. In particular, radio observations with the Very Large Array at 5$\,$GHz were performed by \citet{welch1988} who provided upper limits for the mass-loss rates of the order of $10^{-9}$ to $10^{-7}$~M$_\odot$/yr assuming ionized wind emission. Observations of radio spectral lines were attempted, too. \cite{Matthews2012,Matthews2016} observed the H\,I 21 cm transition with the VLA for different Cepheids. Recently, the first radio detection of continuum emission associated with a Cepheid ($\delta$~Cep) was reported based on VLA observations at 10-15\,GHz \citep{Matthews2020,Matthews2023}. These authors discuss different potential mechanisms responsible for the radio emission, including free-free emission from either a chromosphere, an optically thick ionized shell, or an ionized wind. However, obtaining the spectral index is essential to differentiate these scenarios.

In this letter, we explore the first millimeter-wave observation of the bright Cepheid $\ell$~Car with the Atacama Large Millimeter/submillimeter Array \citep[ALMA,][]{ALMA2009} in order to investigate the presence and the characteristics of the ionized gas emission.
The advantages of observing at higher frequencies are twofold. First, using higher frequencies increases the spatial resolution, which helps mitigate confusion from background sources and potential companions, facilitating the study of regions at sub-arcsecond scales. Second, higher frequencies allow the observation of free-free emission from the stellar wind originating from the inner regions of the circumstellar gas, as the opacity decreases with increasing frequency. Finally, ALMA also allows us to investigate hydrogen recombination lines \citep{Peters2012}. We present the ALMA observations and reduction in Sect.~\ref{sect:observation} and the subsequent results in Sect.~\ref{sect:analysis}, in particular concerning the continuum, the spectral index and the RRL. We conclude in Sect.~\ref{sect:conclusion}.

\section{Observations of $\ell$~Car}\label{sect:observation}

We observed $\ell$~Car with the ALMA main array on 16 December 2023, 5 May, 1 July, 13 July, and 2 October 2024. Observations were conducted in two spectral setups in Band 6. Both setups used the continuum (frequency division mode, FDM) configuration of the correlator with a coarse spectral resolution of 7.8 MHz which is the result of averaging 16 channels before the data were stored. The first setup, {\it Setup~1}, effectively covers simultaneously the spectral ranges 211.6--215.3 and 227.6--231.4\,GHz at a channel binning of about 10.7\,\kms. The second spectral setup, {\it \textit{Setup~2}}, covers higher frequencies, 255.1--258.9 and 270.9--274.6 GHz, at a channel binning of about 8.8\,\kms. The spectral setup was designed not only to provide us with a good estimate of the continuum spectral index, but also to cover a few key spectral lines; for instance, the CO 2--1 and H29$\alpha$ lines at 230.5 and 256.4 GHz, respectively. The CO molecules can form in the Cepheid atmospheres which are metallic enough and have temperature of about 5000\,K \citep{scowcroft2016}. The project was executed at ALMA without a specific requirement on the array configuration, and, as a consequence, our target was observed 3 times in {\it Setup~1} (although two of these were very close in time), and 3 times in {\it \textit{Setup~2}} with different baseline lengths. Details of these observations can be found in Table~\ref{tab:table-obs1}. The visibilities were calibrated by the observatory using the default pipeline in CASA \citep{CASA}. Self-calibration in phases was attempted but did not improve the image quality owing to insufficient signal-to-noise ratio in individual tracks. All imaging was done in CASA 6.6. using {\tt tclean} procedures. Most of the analysis presented in this paper was done on images produced with natural weighting of visibilities, which optimizes the sensitivity of maps at the cost of spatial resolution.


\begin{table*}
\caption{Summary of measurements and derived quantities.}\label{tab:measurements}
\centering\footnotesize
\begin{tabular}{l|cccccc}             
\hline                                         
  & Beam &$S_{\nu}$    & $S_{\nu}$ max     &RMS         &  $\alpha$ &$S_{\nu}$ (BB) \\
  & (\arcsec) &(mJy)     & (mJy/beam)       &   ($\mu$Jy/beam)    & & (mJy) \\
\hline
\textit{Setup~1}        & 0.70$\times$0.59  &3.306     &  3.19       &    10.4      &$+1.45\pm0.16$ &1.23 \\
\textit{Setup~2}         &0.37$\times$0.30  &3.992    & 3.37	&    9.60      & $+1.31\pm 0.09$&1.77 \\
\textit{Setup~1+2}       & 0.46$\times$0.39  & 3.689  &    3.35   &	  7.06   & $+1.26\pm 0.06$&1.50 \\
\hline
\end{tabular}
\tablefoot{All values for maps obtained at natural weighting. Flux density was calculated by a Gaussian fit to continuum images. Max represents the value of maximum pixel (central beam). Stellar photospheric component is not resolved and total fluxes are given. The uncertainty in $\alpha$ represents 1$\sigma$ confidence and does not include the uncertainty in absolute flux. $S_{\nu}$ (BB) is the flux density of the photosphere assuming black-body emission with T$_\mathrm{eff}$ and $\theta_\mathrm{UD}$ from Table~\ref{tab:table-obs1} and derived using Eq.~\ref{eq:Sv}. $S_{\nu}$ (BB) for \textit{Setup 2} is the mean of the three epochs, \textit{Setup 1+2} is the mean of the two setups.}
\end{table*}

\section{Results and discussion}\label{sect:analysis}
Our ALMA observations result in a single unresolved point source at the position of $\ell$~Car with a beam size of about 0\farcs6 and 0\farcs5, respectively, for \textit{Setup 1} and 2 (see Fig.~\ref{fig:setup1_2}). This confirms the absence of any companion of $\ell$~Car brighter than about 35\,$\mu$Jy (5$\sigma$) in Band~6 and within 15\arcsec\ from the star. At uniform weighting of the visibilities which give effective resolution of 85 mas, the map shows a single point source. This is in line with earlier interferometric and radial velocity observations which excluded the possibility of a companion \citep{anderson16lcar}, and is consistent with the analysis of \textit{Gaia} proper motions \citep{Kervella2019}. Flux densities for each Setup and for all continuum data combined are presented in Table~\ref{tab:measurements}, with the flux density of $3.689\pm0.024$\,mJy representing the midpoint of the covered frequencies. We inspected possible variability of the radio emission, in particular between the observations of the second setup which are taken at different phases of the pulsation cycle. We do not find variability between these three epochs above our measurement uncertainties (see Appendix~\ref{appendix-fluxes}).

In order to compare the measured flux density with the stellar continuum, we derived the predicted flux assuming the emission follows the Rayleigh-Jeans regime of a black-body. Indeed, we can estimate the flux of $\ell$~Car at a given frequency knowing its effective temperature and angular diameter. Hereafter we assume that the millimeter continuum is formed at a similar photospheric radius as the optical and the IR. However we note that the millimeter continuum is formed slightly higher in the photosphere, because the atmosphere becomes optically thick at increasing height, as it is the case of the Sun \citep[see, for example,][]{Wedemeyer2016}. In some extreme cases, the presence of gas in the high atmosphere can also form a radio-photosphere of much larger radius such as the ones observed in the case of AGB stars \citep{reid1997}. The flux density $S_\nu$ in the Rayleigh-Jeans tail for a black-body subtended by a solid angle $\Omega$ follows the convenient form:
\begin{equation}\label{eq:Sv}
    S_\nu =\frac{2kT_\mathrm{eff}\nu^2}{c^2}\Omega\approx5.67.10^{-10}\left(\frac{T_\mathrm{eff}}{1\mathrm{K}}\right)\left( \frac{\nu}{1\mathrm{GHz}}\right)^2 \left(\frac{\theta}{1\mathrm{mas}}\right)^2 \mathrm{mJy}
\end{equation}
which is mostly sensitive to the variation of the limb-darkening (LD) angular diameter $\theta$ as compared to the effective temperature $T_\mathrm{eff}$. 

Fortunately, the uniform disk (UD) angular diameter of $\ell$~Car is known precisely along the pulsation cycle owing to interferometric observations in the $H$-band with PIONIER/VLTI \citep{anderson16lcar} and in the $K$-band with VINCI/VLTI \citep{kervella04a} presented in Fig.~\ref{fig:angular_diameter}. In the following, we will neglect the LD effect by adopting the UD angular diameter. On the contrary the effective temperature of $\ell$ Car is not well characterized along the pulsation cycle. We found only one measurement given by \cite{Luck2018}, with $T_\mathrm{eff}=5253\pm38\,$K measured at pulsation phase $\phi=0.05$. In the following we adopt this measurement for the closest epoch of observations. 
For the effectve temperature at $\phi=0.65$ and 0.26, we adopt interpolated values derived using the parallax-of-pulsation code \texttt{SPIPS} \citep{Merand2015}, see Table \ref{tab:table-obs1}. 

\begin{figure}[]
\centering
\includegraphics[scale=0.58]{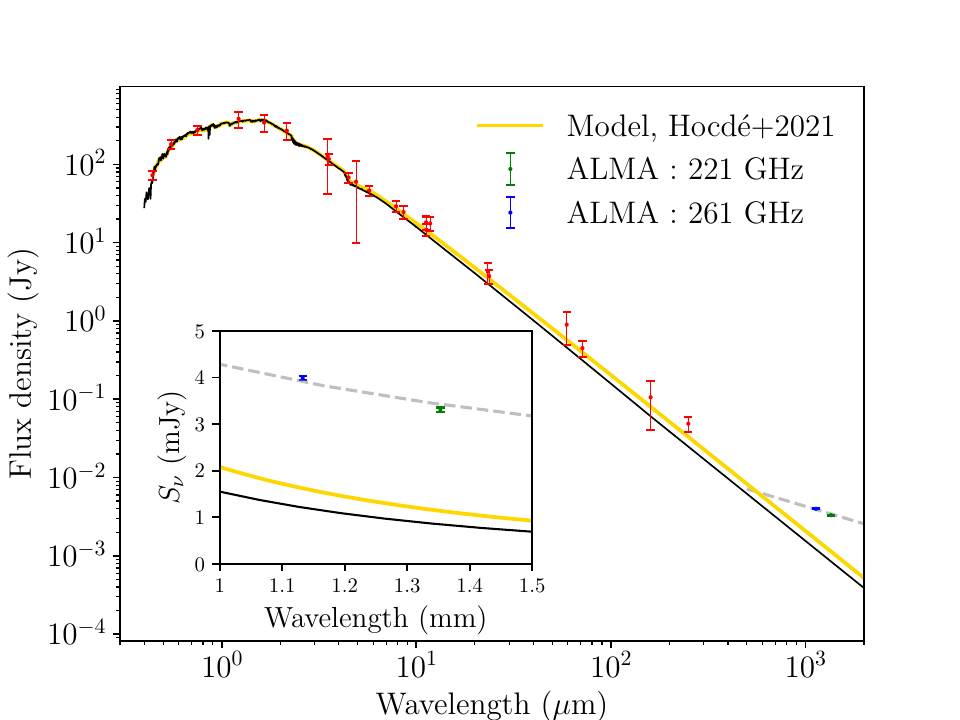}
\caption{\small Spectral energy distribution of $\ell$~Car complemented with ALMA flux density in Band~6 from \textit{Setup~1} (221\,GHz) and \textit{2} (261\,GHz) presented in Table~\ref{tab:table-obs1} and detailed in Sect.~\ref{sect:observation}. 5$\sigma$ errors are displayed on the figure for all the data. Flux density for each band is presented in Table~\ref{tab:flux_density}. The grey dashed line represent a spectral index of $\alpha_\nu=+1.26$ (equivalent to $\alpha_\lambda=-0.74$) as derived in Sect.~\ref{sect:spectral_index}. The SED is derived from an ATLAS9 model (derived up to 10\,$\mu$m and extrapolated beyond) with T$_\mathrm{eff}$ and angular diameter corresponding to the epoch of the \textit{Setup~2} observation, which fits well the optical and near-IR. The envelope model of ionized gas of constant density from \cite{Hocde2021} is shown in yellow. See this paper for a more detailed analysis and a discussion of the impact of free-bound emission in the optical and near-IR bands.}
\label{fig:SED}
\end{figure}

For \textit{Setup~1} at $\phi=0.06$ the measured angular size is $\theta=2.91\pm0.04\,$mas \citep{anderson16lcar} and the temperature measured by \cite{Luck2018} is adopted. Thus, we obtain an expected flux density for \textit{Setup~1} of $S_\nu$$=1.23\pm0.12\,$mJy assuming a conservative 10\% uncertainty because of the black-body assumption. Given the flux density of $S_\nu=3.31\,$mJy measured by ALMA, we obtain a significant excess, 2.7 times brighter than the stellar photospheric continuum. For \textit{Setup~2}, we obtained observations at three different pulsation phases (see Table~\ref{tab:table-obs1}). For the three epochs of $\ell$~Car, we derived an average predicted flux density of $S_\nu=1.77\pm0.18\,$mJy using Eq.~\ref{eq:Sv}. Therefore, the measured flux density is about 2.2 times higher than the stellar continuum. Assuming a maximum $\theta_\mathrm{UD}$ of 3.2\,mas for $\ell$~Car (at $\phi\approx0.4$, see Fig.~\ref{fig:angular_diameter}), the upper limit for the flux density expected is $S_\nu\approx1.9\,$mJy.  We present these measurements in Fig.~\ref{fig:SED} together with photometry obtained in the visible and the IR (see Table \ref{tab:flux_density}) and an ATLAS9 atmosphere model \citep{castelli2003} derived from the effective temperature and the angular diameter corresponding to the pulsation phase of \textit{Setup~2}. From this figure it is clear that the radio emission deviates significantly from Rayleigh-Jeans continuum of the star. We use the measured spectral index
to constrain the mechanism giving rise to the millimeter emission.

\subsection{Spectral index}\label{sect:spectral_index}
For the first time we are able to derive the spectral index of the radio emission of a Cepheid, which is essential for understanding the origin of this emission (see Table~\ref{tab:measurements}). We derived an average spectral index of $+1.26\pm$0.06 (1$\sigma$). 
We calculated the spectral index using CASA's tclean algorithm {\tt mt-mfs}\footnote{\url{https://casadocs.readthedocs.io/en/v6.3.0/notebooks/synthesis_imaging.html}} by applying a second-order ($N$=2) Taylor series. In the resulting maps, the spectral index $\alpha$ shows some variations within the region with strong continuum emission, and thus we take the average value from pixels lying roughly within a contour of 20\% peak flux as the representative $\alpha$; as the associated uncertainty, we take the corresponding average 1$\sigma$ error calculated in tclean (see Appendix \ref{app:spectral_index} for an alternative method). When combining data from two different setups, the results depend quite strongly on the cross-calibration between the basebands. Our flux and bandpass calibration were performed using quasars, for which the observatory claims flux accuracy of about 10\% \citep[but see][]{Francis}. For {\it Setups} 1 and 2 combined, this uncertainty translates to a systematic error in $\alpha$ of about 0.4. We therefore hereafter adopt $\alpha$=+1.26$\pm$0.44 (3$\sigma$).

In the hypothesis that excess flux is caused by free-free emission, the spectral index $\alpha$ ($S_\nu \propto \nu^\alpha$) constrains the physical characteristics of the emission region. For free-free emission from a spherical, symmetrically distributed ionized gas component with constant electron density, the spectral index shifts from $\alpha=-0.1$ in the optically thin regime (observed at frequencies above a critical threshold $\nu>\nu_c$) to $\alpha=+2$ in the optically thick regime at lower frequencies \citep{WrightBarlow1975,panagia1975,Olnon1975}. This is, for example, the model proposed by \cite{Hocde2020a} to explain the IR excess of Cepheids constrained by low-resolution spectra from \textit{Spitzer} which shows optically thick emission in the mid-IR (see the model for $\ell$ Car in Fig.~\ref{fig:SED}). The spectral index at low frequency depends mostly on the density distribution of the ionized gas environment. Hence, the spectral index tends to $+0.6$ at low frequencies for emission in expanding envelopes corresponding to a spherically symmetric ionized wind with a density distribution given as $n_e\propto r^{-2}$. The spectral index calculated in CASA is twice larger than the nominal value of ionized wind emission. This spectral index is also comparable to chromospheric emission of the red supergiant (RSG) Antares in the millimeter regime \citep[$\alpha$=+1.27 at 145\,GHz,][]{OGorman2020} and also Betelgeuse in the centimeter \citep[$\alpha$=+1.33, see][and references therein]{OGorman2015}. In these chromospheric regions of RSGs, the gas is partially optical thick, with a steep gradient in the ionized gas density. Specifically, the spectral index found in the case of $\ell$~Car might be explained if the electron density scales as $n_e \propto r^{-3.0}$ \citep{panagia1975}. However, this spectral index is below the one derived in the model developed for $\ell$ Car and constrained by IR data \citep{Hocde2021} (see Fig.~\ref{fig:SED}). This shows the importance of taking into account the density distribution when modeling the ionized gas emission. It is also not excluded that lower frequencies could probe the wind-dominated emission regime with $\nu^{+0.6}$, while observations at $\sim$200$\,$GHz are most sensitive to the inner regions of the gas envelope. Therefore, future observations at both higher and lower frequencies will help to probe the structure of the ionized gas environment of $\ell$ Car.

\subsection{Radio Recombination Line H29$\alpha$}\label{sect:RRL}
Emission lines from Rydberg transition of hydrogen are produced from radiative recombination at high principal quantum numbers. They are an excellent tracers of ionized gas \citep{BookRRL2002}. For example, emission from H30$\alpha$ was recently detected close to the surface of Betelgeuse at about 230$\,$GHz \citep{Dent2024}. These authors proposed that periodic shocks might be the origin of the emission in the atmosphere.  

Within our \textit{Setup~2} observations we found a spectral feature which we identify as H29$\alpha$, that is a radio recombination line of Rydberg-like transition from principal quantum number $n$=29 of hydrogen. The emission coincides spatially with the continuum emission of $\ell$~Car. At the reference frequency of 256.302035\,GHz, the emission is seen from --54 to 29\,\kms\ (Local Standard of Rest, LSR). Due to a coarse spectral sampling (8.8\,\kms) and a modest signal-to-noise ratio (SNR), this range probably overestimates the spread of the emission. The velocity-integrated emission has a peak at 71.5 mJy/beam \kms\ on maps (see Fig.~\ref{fig:imageh29}) with rms noise of 6.63 mJy/beam \kms.  In the general case, the RRL profile results from the convolution of a Gaussian component accounting for thermal and microturbulent broadening, and a Lorentzian component due to electron pressure broadening, as outlined by \cite{Brocklehurst1972}. Hereafter we assume only a Gaussian profile for simplicity, as the low SNR prevents more detailed modeling.
A Gaussian fit to the feature yields a full width at half power (FWHM) of 55.3$\pm$7.5\,\kms\ and a line center at --6.2$\pm$3.2\,\kms\ (1$\sigma$, LSR). For comparison, \cite{Dent2024} measured a FWHM of $\approx$42\,\kms\ for H30$\alpha$ from the chromosphere of Betelgeuse. The central velocity is equivalent to 7.2\,\kms\ in the Solar System barycenter frame, and thus is within the 3$\sigma$ errors consistent with the catalog systemic velocity of $\ell$~Car of 3.4\,\kms\ in the same rest frame \citep{Anderson2024}.
There is no doubt that the H$\alpha$ emission originates from $\ell$~Car.  The current data do not allow us to conclusively determine the symmetry of the H29$\alpha$ line, which can be useful to probe local thermodynamical equilibrium conditions \citep{Peters2012}. Data at a better SNR and spectral resolution are necessary to verify it.

The flux of $\ell$ Car's H29$\alpha$ emission integrated over the spectral profile is 72.2\,mJy \kms\ or on average 0.9\,mJy per channel in the $\sim$80 \kms\ range occupied by the H$\alpha$ emission. This line flux is approximately equivalent to 35\% of the continuum flux level at the same frequencies (3.5 mJy). The continuum flux is a combination of the photospheric component, estimated at this frequency to be 1.5\,mJy from Eq\,(\ref{eq:Sv}), and the atmospheric or circumstellar component, which we calculate simply as the total measured continuum flux minus the photospheric component, 3.5--1.5=2.0\,mJy (cf. Table~\ref{tab:measurements}). The photospheric contribution was calculated as the blackbody radiation of a star with an effective temperature of 4850\,K, and an angular diameter of 2.9\,mas, representing the average properties of $\ell$~Car. Our calculation therefore leads to a flux ratio of line to non-photospheric continuum of about 0.45. This can be used to calculate the electron temperature of the plasma following the transfer equation for RRLs. With equation 2.124 in \citet{BookRRL2002}, line width of 63 MHz, and an oscillator strength of 5.8174132 \citep{Goldwire1968}, we find that an electron temperature of 23\,500\,K best explains the observed ratio, but the uncertainty in this value is very large, roughly 10\,000 K (propagated from error in the flux of the emission line), due to a very low SNR and uncertain intrinsic line width. Also, the formula assumes optically thin emission for both components. Considering these limitations, we cannot exclude that the electron temperature is considerably higher than 23\,500\,K. On the other hand, assuming optically thick gas we can estimate a lower limit for the mean brightness temperature using Eq.~\ref{eq:Sv}. We find a lower limit of 14\,000\,K and 11,000\,K for \textit{Setup 1} and \textit{2} respectively, assuming the star angular diameter, which rapidly drop off as $\theta^{-2}$ for a larger source.

\section{Conclusions}\label{sect:conclusion}
In this letter, we presented ALMA observations of continuum and H29$\alpha$ emission for the long-period Cepheid $\ell$ Car. The presence of a ionized circumstellar gas is inferred from the following observational results
\begin{enumerate}
    \item We measured continuum emission of about 3.5\,mJy at 200\,GHz (1mm), which is about 2.5 times brighter than the stellar photospheric continuum at these frequencies.
    \item We derived a mean spectral index ($S_\nu\propto \nu^\alpha)$ of $\alpha=+1.26\pm$0.44 indicating partially optically thick ionized gas emission, similar to chromospheric emission from RSGs.
    \item We detected for the first time a RRL of H29$\alpha$ centered on the $\ell$~Car restframe and smaller than about 0\farcs2 ($\approx$100\,au, $\approx$140\,$R_\star$), which is direct evidence for the presence of ionized gas physically connected to the Cepheid. We derived a temperature of the gas of the order of few $10^4\,$K.
    \item We do not find significant evidence of variability among our different epoch of observations.
\end{enumerate}

Hence, these observations reveal the emission of ionized gas connected to the star which is distinct from the photospheric emission. The spectral index derived is higher than expected from free-free emission of partially ionized wind, which prevents us from estimating a mass-loss rate. However, we cannot exclude that such emission can be detected at lower frequency. Our results suggest that the free-free emission originates from the hot gas of the star chromosphere, in line with previous emission lines from the literature for this star \citep{kraft57,schmidt1982I,love1994}. However, the millimeter emission is not explained by the current model of the shell of ionized gas from \cite{Hocde2020a,Hocde2021}, which suggests that this model could be improved by inclusion of density gradient and temperature. Interestingly, we find no significant evidence of variability of the emission in contrast to the results of \cite{Matthews2023} for $\delta$~Cep, which suggests that the free-free emission of $\ell$ Car is not directly linked to the pulsation. The temperature of the gas estimated, of the order of $10^4\,$K, is also not compatible with X-ray emission. The fact that the measurements of the spectral index and RRL of H29$\alpha$ are similar to what is found in RSGs of much lower effective temperature and with very active atmospheres also suggests that convection plays an active role to heat up the gas, as previously noted by \cite{sasselov94a,sasselov94c}.
Finally, given that $\ell$~Car is a nearby analog to the long-period Cepheids observed by the JWST, it is likely that other extragalactic Cepheids possess similar ionized gas emissions, too. As \cite{Neilson2009,Neilson2010} suggested that dust emission could impact the slope and the zero-point of the PL relation in different bands, the question remains whether ionized gas emission could affect the PL relation in the optical and the near-IR.  Further observations are essential to reconstruct a complete radio SED, enabling us to better determine the temperature profile and physical properties of the ionized gas and for a larger sample of Cepheids.

\begin{acknowledgements}
We gratefully thank the referee for valuable comments and suggestions. This paper makes use of the following ALMA data: ADS/JAO.ALMA\#2023.1.00196.S and ADS/JAO.ALMA\#2024.1.00315.S. ALMA is a partnership of ESO (representing its member states), NSF (USA) and NINS (Japan), together with NRC (Canada), NSTC and ASIAA (Taiwan), and KASI (Republic of Korea), in cooperation with the Republic of Chile. The Joint ALMA Observatory is operated by ESO, AUI/NRAO and NAOJ. The research leading to these results has received funding from the European Research Council (ERC) under the European Union’s Horizon 2020 research and innovation programme (grant agreements No 695099 and No 951549). This work received the funding from the Polish-French Marie Skłodowska-Curie and Pierre Curie Science Prize awarded by the Foundation for Polish Science and the Polish Ministry of Science and Higher Education grant agreement 2024/WK/02. The authors acknowledge the support of the French Agence Nationale de la Recherche (ANR) under grant
ANR-23-CE31-0009-01 (Unlock-pfactor).
  This research made use of the SIMBAD and VIZIER (Available at \url{http://cdsweb.u- strasbg.fr/}) databases at CDS, Strasbourg (France) and the electronic bibliography maintained by the NASA/ADS system. 
This research also made use of Astropy, a  community-developed core Python package for Astronomy \citep{astropy2018}. This research has benefited from the help of SUV, the VLTI user support  service of the Jean-Marie Mariotti Center (\url{http://www.jmmc.fr/suv.htm}). This research has also made use of the Jean-Marie Mariotti Center \texttt{Aspro}
service (Available at \url{http://www.jmmc.fr/aspro}).
\end{acknowledgements}
\bibliographystyle{aa}  
\bibliography{bibtex_vh} 

\begin{appendix} 
\onecolumn
\section{Observational material}
\begin{figure*}[h!] 
\centering
\begin{subfigure}{0.41\textwidth}
\includegraphics[width=\linewidth]{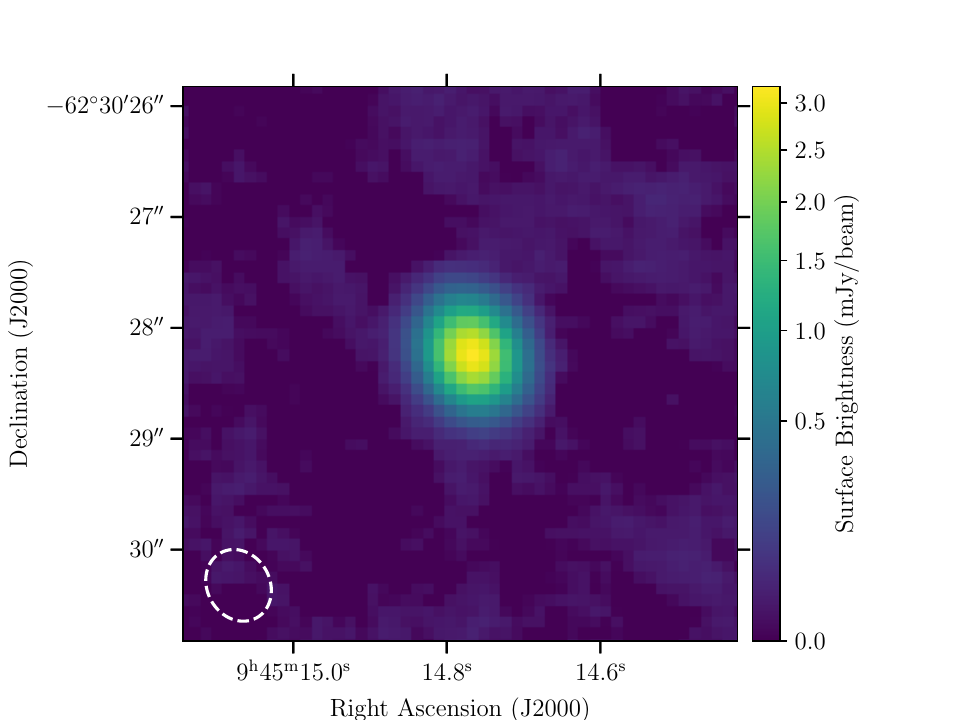}
\caption{}  \label{fig:Setup1}
\end{subfigure}
\begin{subfigure}{0.41\textwidth}
\includegraphics[width=\linewidth]{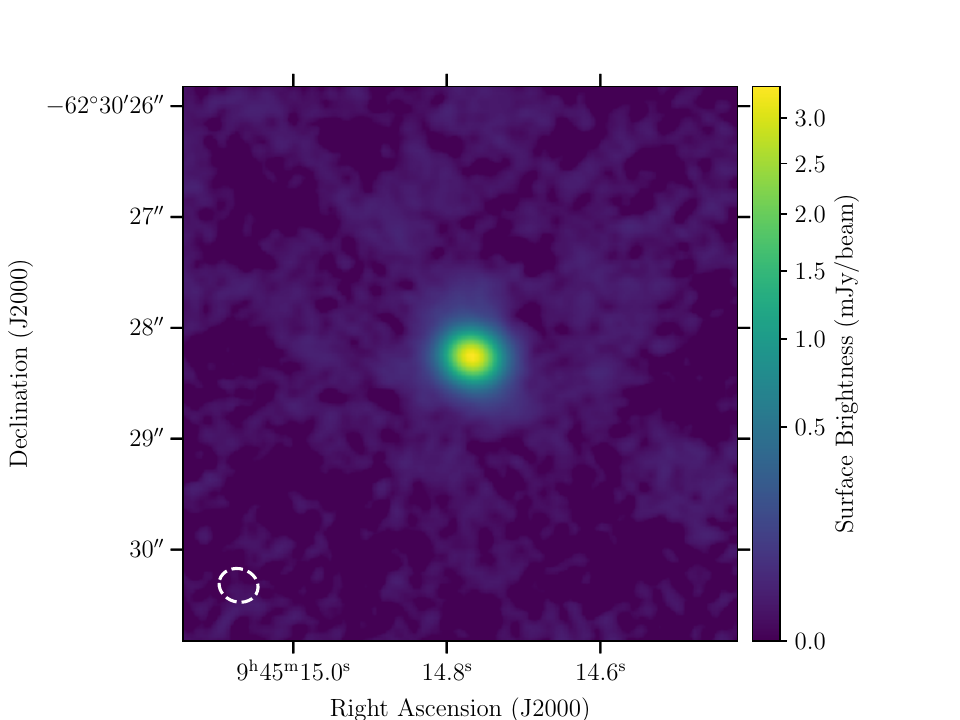}
\caption{}  \label{fig:Setup2a}
\end{subfigure}
\caption{\small ALMA continuum observation of $\ell$~Car in (left) \textit{Setup~1} (221.61 GHz)  and (right) \textit{Setup~2} (264.86 GHz) displayed on a $5\arcsec \times 5\arcsec$ field-of-view. The beam sizes for the two Setups are given on the bottom left corner (see also Table \ref{tab:flux_density}).}\label{fig:setup1_2}
\end{figure*}

\begin{figure*}[h!]
\centering
\begin{subfigure}{0.41\textwidth}
\includegraphics[width=\linewidth]{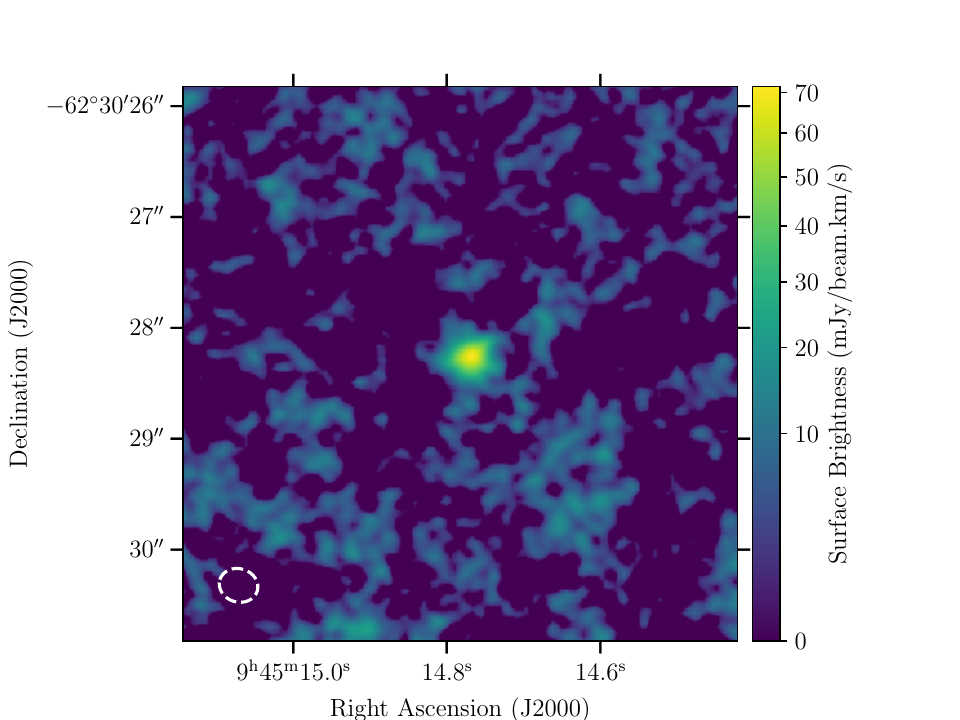}
\caption{}  \label{fig:imageh29}
\end{subfigure}
\begin{subfigure}{0.37\textwidth}
\includegraphics[width=\linewidth]{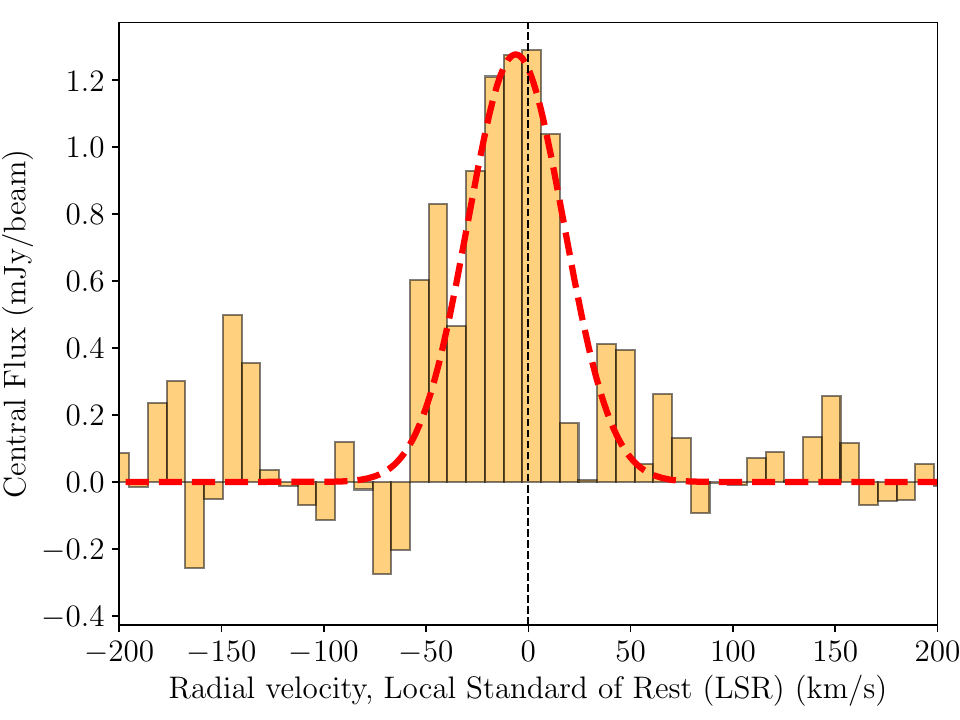}
\caption{}  \label{fig:H29line}
\end{subfigure}
\caption{\small Detection of the radio recombination line H29$\alpha$ toward $\ell$ Car (see Sect.~\ref{sect:RRL}). The left map presents the total emission integrated over the line profile, while the right panel presents the spectrum extracted at the pixel with maximum emission. Continuum emission was subtracted from the visibilities before imaging of the line emission. The histogram presents the observations, while the red dashed line present a Gaussian fit to the line profile.}\label{fig:H29plots}
\end{figure*}

The observations hereafter are introduced in Sect.~\ref{sect:observation}. Figs.~\ref{fig:setup1_2} and \ref{fig:H29plots} present the ALMA continuum observations and the radio recombination line H29$\alpha$ respectively.
\begin{table}[h!]
\caption{Journal of ALMA observations together with $\ell$ Car temperature and angular diameter.}        
\label{tab:table-obs1}     
\centering                        
\begin{tabular}{c c c c c c l|c c c}       
\hline                                                               
     &  Date & Date  & $t_{\rm on}$	 &PWV   &Baselines  & $N_{\rm ant}$ & $\phi$ & T$_\mathrm{eff}$ & $\theta_\mathrm{UD}$ \\
           & obs. UTC	 & (MJD) & (min)  &(mm)  &(m)	&      & &(K) &(mas) \\
\hline	        		   		         	        
   \textit{Setup~1}       & 2024-05-05 01:04:51 &60435.04 &  38   &1.19  &15 -- 99  & 42     &	0.02&5253$\pm$38  & $2.91\pm0.04$ \\
     (221.61 GHz)             & 2023-12-16 09:05:51 & 60294.38 &  38   &3.26  &15 -- 2516  & 47   &	0.06 & 5253$\pm$38 & $2.91\pm0.04$\\
                  & 2023-12-16 08:08:02 & 60294.34 & 38   &2.98  &15 -- 2516  & 47   &	0.06 &5253$\pm$38 &$2.91\pm0.04$ \\				        
                  \hline	          		   		         	      
\textit{Setup~2}      & 2024-07-13 21:46:16 &60504.91 &  44   &0.34  &15 -- 2516  & 43   &	0.98& 5253$\pm$38 & $2.91\pm0.04$ \\
   (264.86 GHz)       & 2024-07-01 20:23:21 & 60492.85 & 44   &0.31  &37 -- 2516  & 43   &	0.65 & 4586$\pm$100 &$2.99\pm0.04$\\
           & 2024-10-02 13:50:51 & 60585.56 & 41 & 0.94 & 15 -- 499 & 42 & 0.26 & 4785$\pm100$ & $3.18\pm0.04$ \\

\hline
\end{tabular}
\tablefoot{Journal of observations presented in Sect.~\ref{sect:observation} for \textit{Setup 1} and 2. $t_{\rm on}$ is time on source. PWV is the average precipitable water vapour during the observations. Baseline lengths represent datasets before pipeline flagging. $N_{\rm ant}$ is the number of unflagged antennas. For each epoch, we derive the pulsation phase using recent pulsation period and reference epoch at maximum light as measured by \textit{Gaia}, $P_0=35.558\,$day and $T_0=56807.202$ \citep{gaia2023} and with a rate of period change from \cite{Csorney2022}. For each pulsation phase $\phi$, the UD angular diameter $\theta_\mathrm{UD}$ are retrieved from the literature, while the effective temperature $T_\mathrm{eff}$ are retrieved from the literature and interpolated (see Sect.~\ref{sect:analysis}).}
\end{table}


\twocolumn
\clearpage
\section{Photometry and angular diameter measurements}

\begin{figure}[h]
\centering
\includegraphics[scale=0.5]{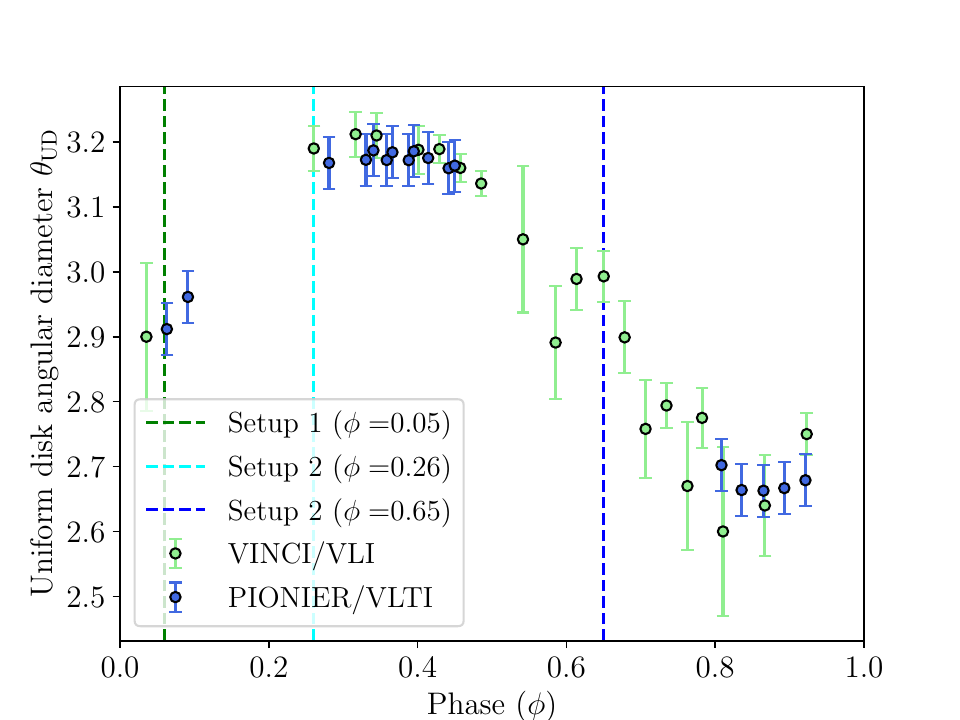}
\caption{\small Uniform disk angular diameter variation of $\ell$~Car obtained from VINCI/VLTI and PIONIER/VLTI measurements \citep{kervella04a,anderson16lcar}. The dashed lines represent the three pulsation phases of observation at $\approx$0.05, 0.26 and 0.65 for \textit{Setup 1} and 2 (see Table \ref{tab:measurements}). }\label{fig:angular_diameter}
\end{figure}

\begin{table}[h]
    \centering
    \begin{tabular}{l|c|c|c}
        \hline
        \hline
        Inst.&$\lambda_\mathbf{eff}$  & Flux density & References \\
        \textit{System}&($\mu$m) & (Jy) &\\
        \hline
        \textit{Johnson}&0.44  & 72.6 $\pm$ 2.0 &F07  \\
         \textit{Johnson}&0.55  & 181 $\pm$ 5.1  &F07   \\
        \textit{Cousins}&0.75  & 274 $\pm$ 7.7  &F07   \\
        \textit{2MASS}&1.22  & 382 $\pm$ 18   &F07   \\
        \textit{2MASS}&1.65  & 342 $\pm$ 16   &F07   \\
        \textit{2MASS}&2.16  & 269 $\pm$ 13   &F07   \\
        COBE&3.50  & 127 $\pm$ 17   & S04 \\
        IRAC&3.55  & 117 $\pm$ 4    & K09 \\
        IRAC&4.49  & 68 $\pm$ 2     &  K09\\
        COBE&4.90  & 60 $\pm$ 10    & S04 \\
        IRAC&5.73  & 46.3 $\pm$ 1.4 & K09 \\
        IRAC&7.87  & 29.1 $\pm$ 0.9 &  K09\\
        VISIR&8.60  & 24.6 $\pm$ 0.9 &  K09\\
        VISIR&11.22 & 18.1 $\pm$ 0.7 &  K09\\
        COBE&11.25 & 14.6 $\pm$ 0.5 &  S04\\
        IRAS&11.79 & 13.9 $\pm$ 0.3 &  I19\\
        IRAS&23.34 & 4.31 $\pm$ 0.26 & I19\\
        MIPS&23.68 & 3.73 $\pm$ 0.15 & K09\\
        IRAS & 59.32& 0.896	$\pm$ 0.0801 & I19 \\
        MIPS&71.42 & 0.45 $\pm$ 0.02 & K09 \\
        PACS & 160	& 0.106 $\pm$ 0.013 & M24\\
         SPIRE &    250 &	0.0488 $\pm$ 0.0021 & H24\\
         SPIRE & 363 &	0.0288 $\pm$ 0.0023 & H24\\
        \hline
    \end{tabular}
    \caption{Flux density in Janskys (Jy) at different wavelengths with associated uncertainties. The SED is displayed in Fig.~\ref{fig:SED} complemented with ALMA observations. I19: \cite{Neugebauer2019}, S04: \cite{Smith2004}, F07: \cite{Fouqué2007}, K09: \cite{Kervella2009}, M24: \cite{Marton2024}, H24: \cite{Herschel2024}.}
    \label{tab:flux_density}
\end{table}

\section{Spectral index fit}\label{app:spectral_index}
\begin{figure}[h!]
\centering
\includegraphics[scale=0.7]{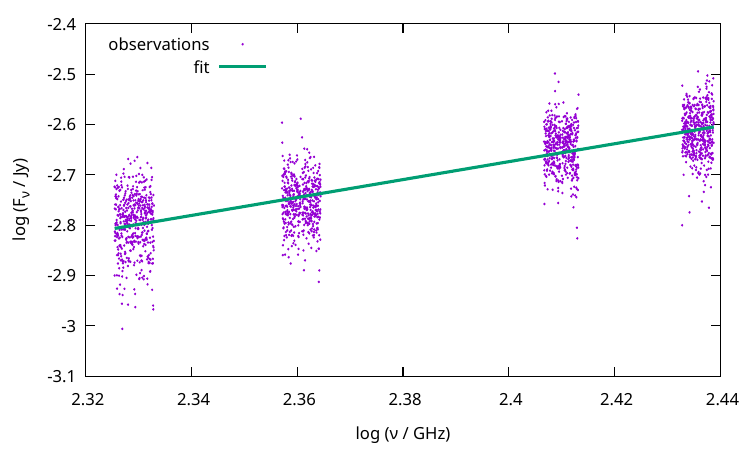}
\caption{\small Fit to combined spectra from Setups 1 and 2. Spectra were extracted from maps obtained at natural weighting and with the same
synthesized beam. The extraction aperture was defined by the contour at 20\% of maximum flux. }\label{fig:spfit}
\end{figure}

As an alternative method of deriving $\alpha$ presented in Sect.~\ref{sect:spectral_index}, we generated a full spectrum of the map area within the 20\% contour and fitted a line to the logarithmic version of the spectrum ($\log \lambda$ vs $\log F_{\nu}$), as shown in Fig. \ref{fig:spfit}. This resulted in $\alpha$=+1.78$\pm$0.03 (1$\sigma$), a value higher than that obtained in CASA in Sect.~\ref{sect:spectral_index}. However, since the simple linear fit is more prone to calibration errors between the different frequency setups of the correlator \citep[cf.][]{Francis}, we favor the result obtained in CASA.

\section{Fluxes in individual epochs}\label{appendix-fluxes}
Flux densities were measured for each individual observation on restored images and using a Gaussian fit. They are presented in Table~\ref{tab:table-obs2} along with the associated 1$\sigma$ errors derived from the statistical noise in the data and to the goodness of the fit \footnote{\url{https://casadocs.readthedocs.io/en/latest/api/tt/casatasks.analysis.imfit.html}}. The fluxes measured for the last epoch in {\it Setup~2}, on 2 Oct 2024, are over 30\% higher than in both earlier epochs. Although this discrepancy exceeds the 5$\sigma$ statistical uncertainties and 10\% calibration errors from ALMA documentation, all measurements are consistent within the more realistic statistical errors from practical applications, as shown by \citep{Francis}.

\begin{table}[h]
\caption{Flux measurements for individual observations.}         
\label{tab:table-obs2}     
\centering                        
\begin{tabular}{c c c c c}       
\hline                                                               
     &  Date    & Flux & 1$\sigma$ error & Beam size \\
     & obs. UTC& (mJy)         & (mJy)  &(\arcsec) \\
\hline	        		   		         	        
     & 2023-12-16 09:05:51 & 3.506 & 0.038 & 0.307$\times$0.270 \\
     & 2023-12-16 08:08:02 & 3.562 & 0.044 & 0.317$\times$0.301 \\
     & 2024-05-05 01:04:51 & 3.213 & 0.066 & 1.136$\times$0.954 \\
\hline	        		   		         	        
     & 2024-07-13 21:46:16 & 3.265 & 0.021 & 0.213$\times$0.181 \\
     & 2024-07-01 20:23:21 & 3.821 & 0.040 & 0.228$\times$0.194 \\
     & 2024-10-02 13:50:51 & 4.653 & 0.032 & 0.869$\times$0.773 \\
\hline
\end{tabular}
\tablefoot{Flux is integrated flux density measured by a Gaussian fit to an image. 1$\sigma$ error is the statistical uncertainty of the fit taken from the CASA's {\tt{imfit}} procedure, where the noise rms was specified based on image noise statistics. The uncertainties do not include statistical (calibration) uncertainties.}
\end{table}

\end{appendix}
\end{document}